# Project Lyra: Sending a Spacecraft to 1I/'Oumuamua (former A/2017 U1), the Interstellar Asteroid

Andreas M. Hein[1], Nikolaos Perakis[1], T. Marshall Eubanks[1,2], Adam Hibberd[1], Adam Crowl[1], Kieran Hayward[1], Robert G. Kennedy III[1], Richard Osborne[1]

Contact email: andreas.hein@i4is.org

[1] Initiative for Interstellar Studies, Bone Mill, New Street, Charfield, GL12 8ES, United Kingdom

[2] Asteroid Initiatives LLC

**Abstract**

The first definitely interstellar object 1I/'Oumuamua (previously A/2017 U1) observed in our solar system provides the opportunity to directly study material from other star systems. Can such objects be intercepted? The challenge of reaching the object within a reasonable timeframe is formidable due to its high heliocentric hyperbolic excess velocity of about 26 km/s; much faster than any vehicle yet launched. This paper presents a high-level analysis of potential near-term options for a mission to 1I/'Oumuamua and potential similar objects. Launching a spacecraft to 1I/'Oumuamua in a reasonable timeframe of 5-10 years requires a hyperbolic solar system excess velocity between 33 to 76 km/s for mission durations between 30 to 5 years. Different mission durations and their velocity requirements are explored with respect to the launch date, assuming direct impulsive transfer to the intercept trajectory. For missions using a powered Jupiter flyby combined with a solar Oberth maneuver using solid rocket boosters and Parker Solar Probe heatshield technology, a Falcon Heavy-class launcher would be able to launch a spacecraft of dozens of kilograms towards 1I/'Oumuamua, if launched in 2021. An additional Saturn flyby would allow for the launch of a New Horizons-class spacecraft. Further technology options are outlined, ranging from electric propulsion, and more advanced options such as laser electric propulsion, and solar and laser sails. To maximize science return decelerating the spacecraft at 'Oumuamua is highly desirable, due to the minimal science return from a hyper-velocity encounter. Electric and magnetic sails could be used for this purpose. It is concluded that although reaching the object is challenging, there seem to be feasible options based on current and near-term technology.

## 1. Introduction

On October 19[th] 2017, the University of Hawaii's Pan-STARRS 1 telescope on Haleakala discovered a fast-moving object near the Earth, initially named A/2017 U1, but now designated as 1I/'Oumuamua [1]. It is likely that this object has its origin outside the solar system [2–9], with a velocity at infinity of 26.33 km/s, an eccentricity of 1.20, and an incoming radiant (direction of motion) near the solar apex in the constellation Lyra [10]. Due to the non-observation of a tail in the proximity of the Sun, the object seems to be an asteroid [11]. However, it has been hypothesized that either a cometary tail was present for a brief moment but was not observed [12], as 1I/'Oumuamua was discovered post perihelion, or that an organically rich surface, resulting in an insulating mantle prevented a cometary tail from forming [13]. Such a mantle could be the result of long-term cosmic ray exposure, although Jackson et al. argue against this possibility [14]. The comet hypothesis has been supported by the observation of non-gravitational acceleration, which can be explained by cometary outgassing [15]. However, the question is still far from resolved [16]. Spectroscopic results [12,13,17–21] indicate that the object is reddish, with a distribution similar to Trans-Neptunian objects [18,19,21]. The rapidly changing albedo has also lead to the assumption that the object is rotating and is highly elongated [19,22,23] with estimated dimensions of 230m x 35m x 35m [21]. The axis ratio of $\geq 6.3^{+1.3}_{-1.1}:1$ is larger than for any small solar system body [21]. Its orbital features have been analyzed by [10,24,25]. Hypotheses for its origin range from a star of the Local Association [4] to a more distant origin in the galactic thin disk, with the ejection dating back several billion years [7,20].

Estimates for the abundance of interstellar objects with a similar size are wide-ranging. Feng and Jones [4] estimate an abundance of interstellar objects larger than 100 m as $6.0 \times 10^{-3} AU^{-3}$. The steady state population of interstellar objects with a size of the order of 100 m inside the orbit of Neptune has been estimated as on the order of $10^4$ by [21]. As 1I/'Oumuamua is the nearest macroscopic sample of interstellar material, likely with an isotopic signature distinct from any other object in our solar system, the scientific returns from sampling the object are hard to understate. Detailed in-situ studies of interstellar materials at interstellar distances is likely decades away. For example, Breakthrough Initiatives' Project Starshot, which is developing a laser-propelled interstellar probe along with a beaming infrastructure for example, currently aims at a launch date in the 2040s [26]. Hence, an interesting question is if there is a way to exploit this unique opportunity to study interstellar material by sending a spacecraft to 1I/'Oumuamua to make observations at close range.

The Initiative for Interstellar Studies, i4is, has announced Project Lyra on the 30[th] of October to answer this question. The goal of the project is to assess the feasibility of a mission to 1I/'Oumuamua using current and near-term technology and to propose mission concepts for achieving a fly-by or rendezvous. The challenge is formidable. According to current estimates, 1I/'Oumuamua has a heliocentric hyperbolic excess velocity of 26 km/s. This is considerably faster than any object humanity has ever launched into space. Voyager 1, the fastest object humanity has ever built, has a hyperbolic excess velocity of 16.6 km/s. As 1I/'Oumuamua is already leaving our solar system, any spacecraft launched in the future would need to chase it. However, besides the scientific interest of getting data back from the object, the challenge to reach the object could stretch the current technological envelope of space exploration. Hence, Project Lyra is not only interesting from a scientific point of view but also in terms of the technological challenge it presents. This paper presents some preliminary results for a mission concept to 1I/'Oumuamua.

## 2. Trajectory Analysis

In the following, we first provide a first order trajectory analysis assuming one or more impulsive transfers in the vicinity of the earth and a free trajectory to 1I/'Oumuamua. Subsequently we present results for a more complex trajectory that combines a Jupiter flyby and a solar Oberth maneuver.

### 2.1 Trajectory Analysis without Gravity Assist

Given the hyperbolic excess velocity and its inclination with respect to the solar system ecliptic, the first question to answer is the required velocity increment (DeltaV) to reach the object, a key parameter for designing the propulsion system. The DeltaV for a mission without gravity assist is calculated by determining the transfer hyperbola from Earth orbit with respect to the position of 1I/'Oumuamua at a certain point in time. Using the vis-viva equation (1), the orbital velocity $v$ of a body on a hyperbolic trajectory can be computed.

$$v = \sqrt{\mu \left(\frac{2}{r} - \frac{1}{a}\right)} \quad (1)$$

Where $\mu$ is the standard gravitational parameter, $r$ is the radial distance of the object from the central body, and $a$ the semi-major axis. With a few arithmetic manipulations, the relationship (2) between the orbital velocity $v$, the escape velocity $v_{esc}$ from the Sun, and the hyperbolic excess velocity $v_\infty$ can be obtained.

$$v = \sqrt{v_\infty^2 + \frac{2\mu}{r}} \quad (2)$$

$v_\infty$ can be understood as the velocity at infinity with respect to the Sun. Note that the earth's orbital velocity can be exploited for reducing the required DeltaV. Nevertheless the high inclination of 1I/'Oumuamua relative to the ecliptic requires significant additional DeltaV.

$$\Delta v = v - v_{earth} \cos i \cos \eta = \sqrt{v_\infty^2 + \frac{2\mu}{r}} - v_{earth} \cos i \cos \eta \qquad (3)$$

Where $i$ is the inclination of the trajectory with respect to the ecliptic and $\eta$ the angle between the velocity vector of the trajectory and the velocity vector of the Earth.

Obviously, a slower spacecraft will reach the object later than a faster spacecraft, leading to a trade-off between trip duration and required DeltaV. Furthermore, the earlier the spacecraft is launched the shorter the trip duration, as the object is closer. However, a launch date within the next 5 years is likely to be unrealistic, and even 10 years could be challenging, in case new technologies need to be developed. A third trade-off is between launch date and trip time / characteristic energy $C_3$. The characteristic energy is the square of the hyperbolic excess velocity $v_\infty$. These trade-offs are captured in Figure 1. The figure plots the characteristic energy for the launch with respect to mission duration and launch date. An impulsive propulsion system with a sufficiently short thrust duration is assumed. No planetary or solar fly-by is assumed, only a direct launch towards the object. It can be seen that a minimum $C_3$ exists, which is about 703 km²/s² (26.5 km/s). However, this minimum value rapidly increases when the launch date is moved into the future. At the same time, a larger mission duration leads to a decrease of the required $C_3$ but also implies an encounter with the asteroid at a larger distance from the Sun. A realistic launch date for a probe would be at least 10 years in the future (2027). At that point, the hyperbolic excess velocity with respect to the Earth is already at 37.4km/s (1400 km²/s²) with a mission duration of about 15 years, which makes such an orbital insertion extremely challenging with conventional launches in the absence of a planetary fly-by.

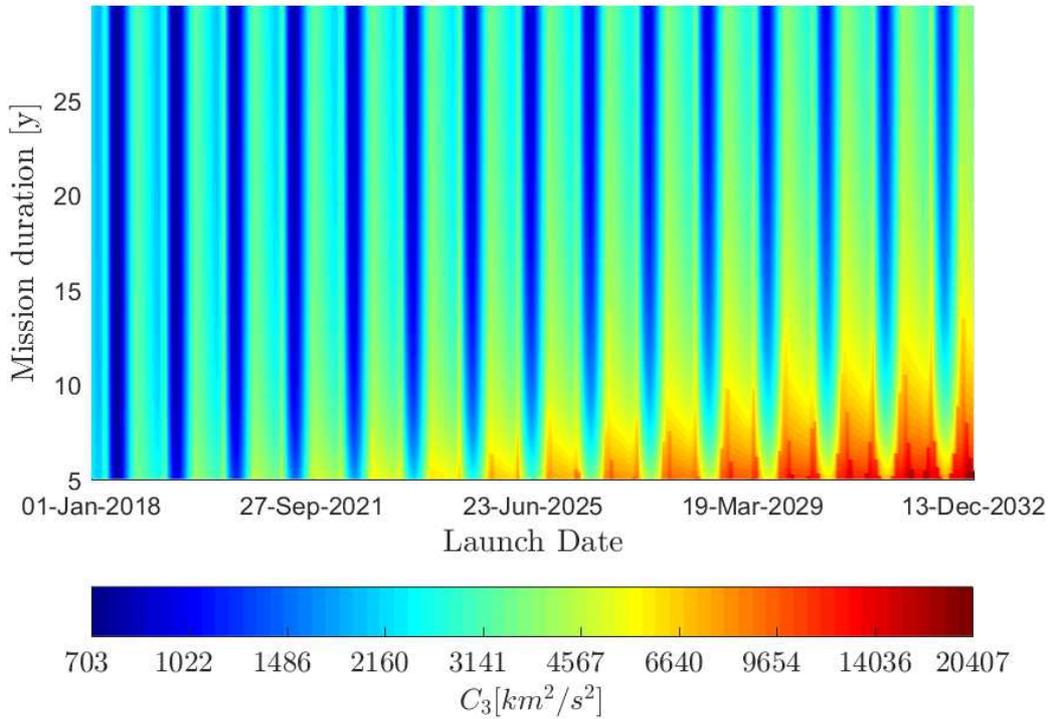

*Figure 1: Characteristic energy C3 with respect to mission duration and launch date.*

Apart from the hyperbolic excess velocity at launch, the excess velocity relative to the asteroid at encounter ($v_{\infty,2}$) has to be taken into account, since it defines the type of mission that is achievable. $v_{\infty,2}$ can be calculated via equation (4), where $v_{\infty,2}$ is the difference between the hyperbolic excess of the spacecraft $v_{\infty,SC}$ and the hyperbolic excess of 1I/'Oumuamua $v_{\infty,1I}$.

$$v_{\infty,2} = v_{\infty,SC} - v_{\infty,1I} \qquad (4)$$

A high excess velocity with respect to the asteroid reduces the flight duration but also reduces the time available for the measurements close to the interstellar object. On the other hand, a low value for $v_{\infty,2}$ could even enable orbital insertion around the asteroid with an impulsive or low thrust maneuver to decelerate the probe. The excess velocity at arrival is plotted in Figure 2 as a function of the launch date and the flight duration. The deformations of the velocity curves is due to the Earth's orbit around the Sun, which results in a more or less favorable position for a launch towards the object. It can be seen that a minimum excess velocity of about 26.75km/s implies a launch in 2018 and a flight duration of over 20 years. Such value for excess velocity does not prohibit an orbital insertion around 'Oumuamua. However, this minimum value rapidly increases for later launch dates. A realistic launch date for a probe would be between 5 to 10 years in the future (2023 to 2027). At that point, the required hyperbolic excess velocity for the mission is between 33 to 76 km/s, for mission durations between 30 to 5 years. These values highly exceed the current chemical and electric propulsion system capabilities for deceleration and orbital insertion, and hence a fly-by would be more reasonable.

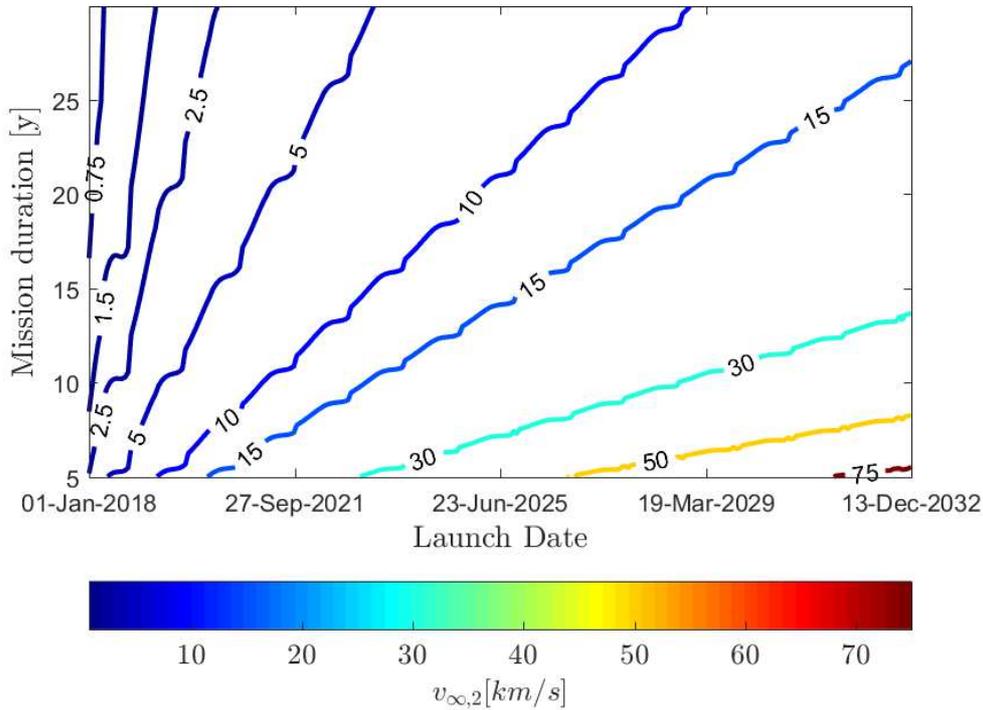

*Figure 2: Encounter velocities with respect to mission duration and launch date*

Figure 3 shows the approximate distance at which the spacecraft passes the object. For a realistic launch date of 2027 or later, the spacecraft flies past the object at a distance between 100 and 200 AU, which is similar to the distance to the Voyager probes today. At such a distance, obviously power and communication becomes an issue and nuclear power sources such as RTGs are required.

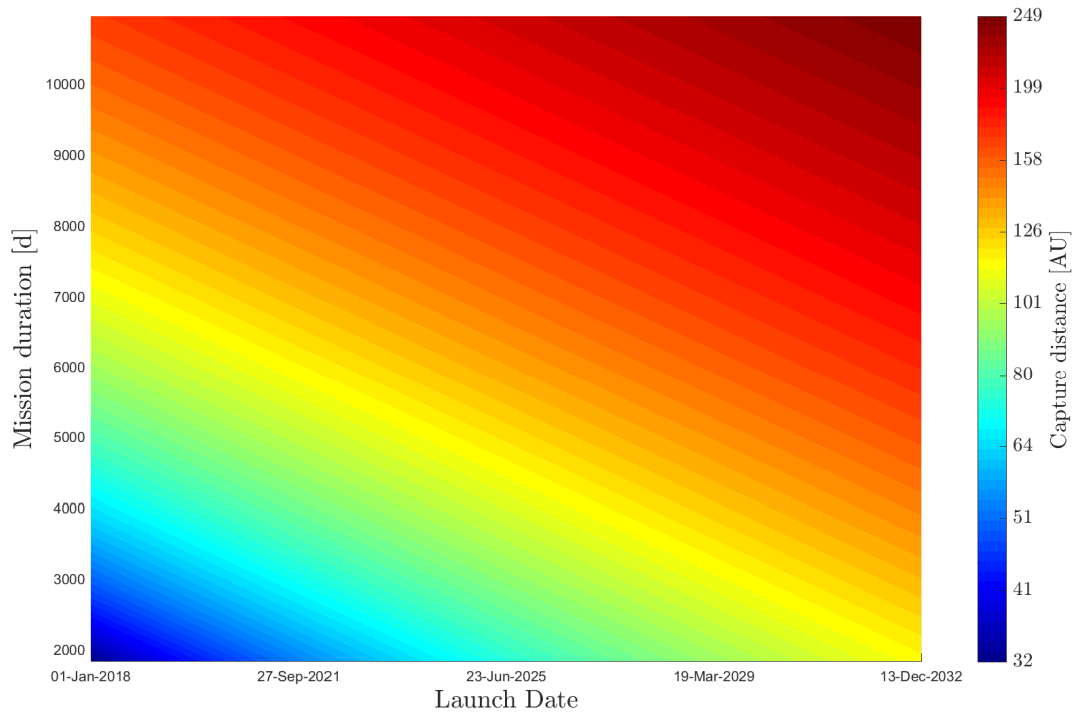

*Figure 3: Launch date versus mission duration. Color code indicates the distance at which the spacecraft passes the object*

Figure 4 shows a sample trajectory with a launch date in 2025. The orbit of Earth can be seen as a tiny ellipse around the Sun (indicated as a black circle) at the bottom right of the figure. The trajectories of the comet and the spacecraft are almost straight lines.

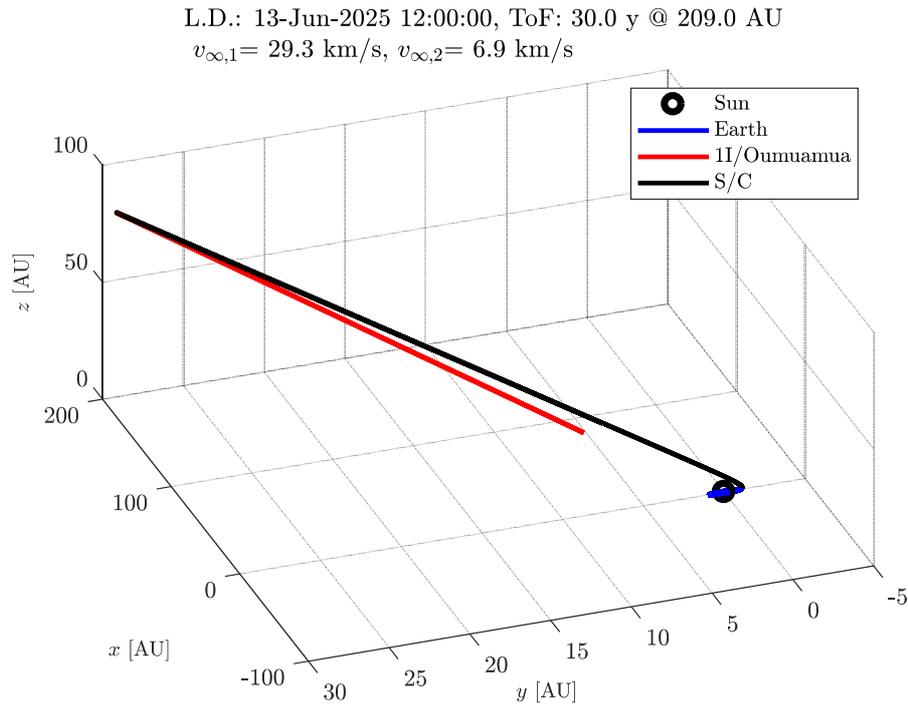

*Figure 4: Sample spacecraft trajectory for a launch in 2025 and an encounter with 1I/'Oumuamua in 2055*

Another proposal is to not necessarily chase 1I/'Oumuamua but to prepare for the next interstellar object to enter our solar system by developing the means to quickly launch a spacecraft towards such an object.

Two scenarios are analyzed: First a mission with short duration of only a year, leading to an encounter only 5.8 AU from the Sun. However the required hyperbolic excess velocity the current launcher capabilities at approximately 20 km/s. Finally, due to the angle of the encounter, a high velocity relative to the asteroid would be expected, amounting to 13.6 km/s, shown in Figure 5.

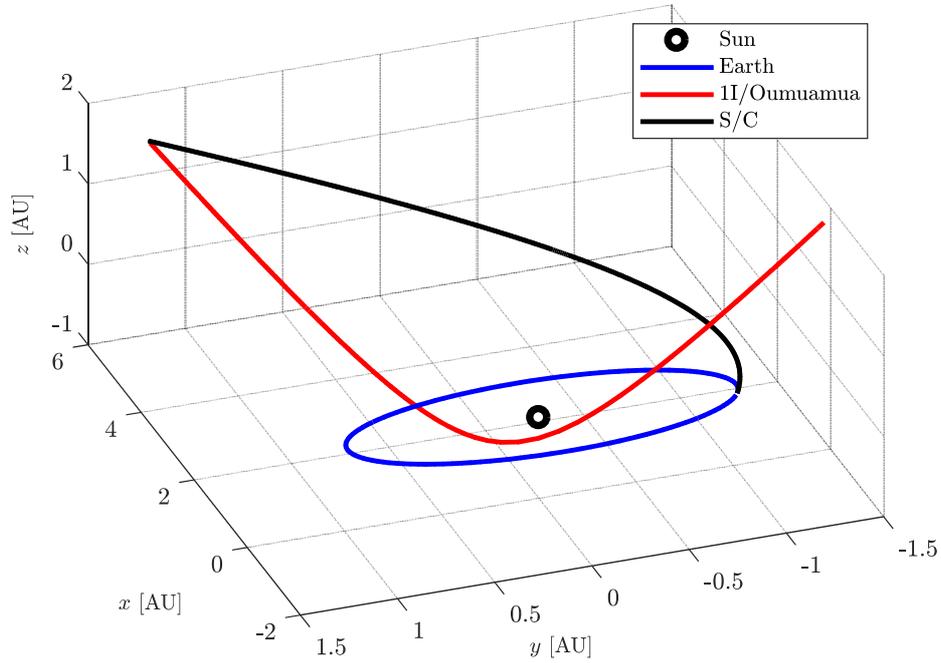

*Figure 5: Trajectory for a launch in 2017 and an encounter in 2018*

A mission on the same launch date but with a duration of 20 years is shown in Figure 6. At encounter, the relative velocity of the spacecraft with respect to the object is relatively low (about 600 m/s for this specific case), which would be an opportunity for a deceleration maneuver.

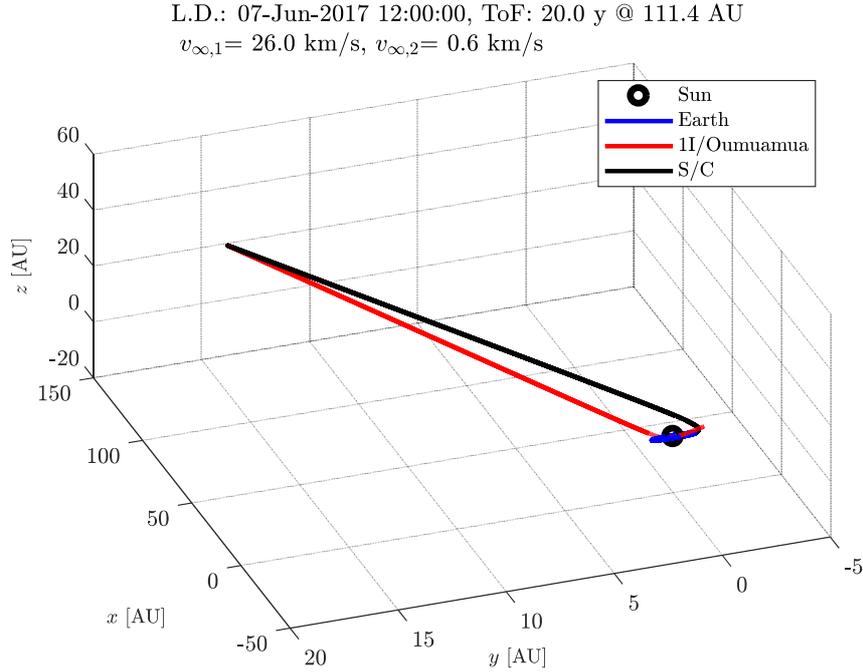

*Figure 6: Trajectory for a launch in 2017 and an encounter in 2037*

To summarize, the difficulty of reaching 1I/'Oumuamua is a function of when to launch, the hyperbolic excess velocity, and the mission duration. Future mission designers would need to find appropriate trade-offs between these parameters. For a realistic launch date in 5 to 10 years, the hyperbolic excess velocity is of the order of 33 to up to 76 km/s with an encounter at a distance far beyond Pluto (50-200AU).

### 2.2 Trajectory Analysis with Solar Oberth Maneuver

To achieve the required hyperbolic excess (at least 30 km/s) for a rendezvous with 1I/'Oumuamua using chemical propulsion systems, a Jupiter flyby is combined with a close solar flyby (down to 3 solar radii). The maneuver is also known under "Oberth Maneuver" [27,28]. The architecture is based on the Keck Institute for Space Studies (KISS) [29] and the Jet Propulsion Laboratory (JPL) [30] interstellar precursor mission studies. In the following, a few results for this mission architecture are presented. Details about the required technologies is provided in section 3.

For calculating the trajectory, the Optimum Interplanetary Trajectory Software (OITS) developed by Adam Hibberd was used. A patched conic approximation is applied, i.e. within the sphere of influence of a celestial body, only its respective gravitational attraction is taken into consideration and the gravitational attraction of other bodies is neglected. The trajectory connecting each pair of celestial bodies is determined by solving the Lambert problem using the Universal Variable Formulation [31]. The resulting non-linear global optimization problem with inequality constraints is solved using the NOMAD solver.

The resulting minimal DeltaV for an eight year flight duration is shown in Figure 7. The DeltaV varies between 18 km/s for a launch date in April to May 2021 up to a value of 52 km/s for a launch in October 2021. A second dashed curve shows the minimum DeltaV for a trajectory without the Jupiter flyby and the solar Oberth maneuver. It can be seen that the DeltaV is always higher, although the difference is only about 5 km/s for a launch in July to September.

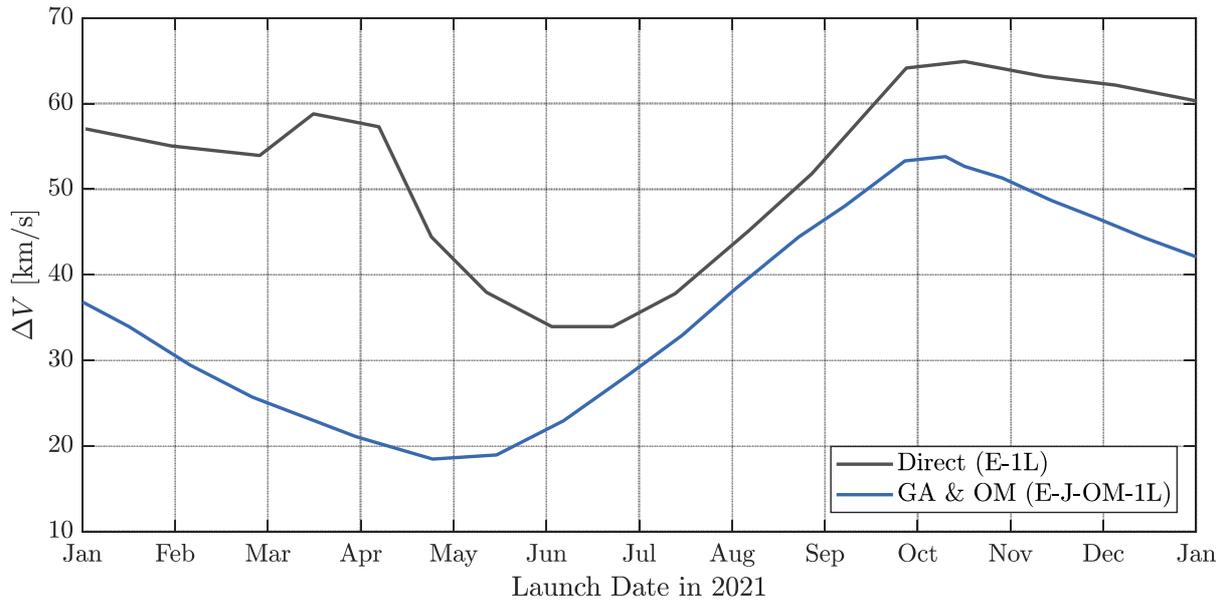

*Figure 7: Minimum DeltaV for combined Jupiter gravity assist (GA) and solar Oberth maneuver (OM) and a launch date in 2021*

Figure 8 shows a visualization of the trajectory with the first leg from Earth to Jupiter, the second leg from Jupiter to the Sun, and the subsequent encounter trajectory with 1I/'Oumuamua with an encounter at 69 AU. The hyperbolic excess velocity is 55.7 km/s.

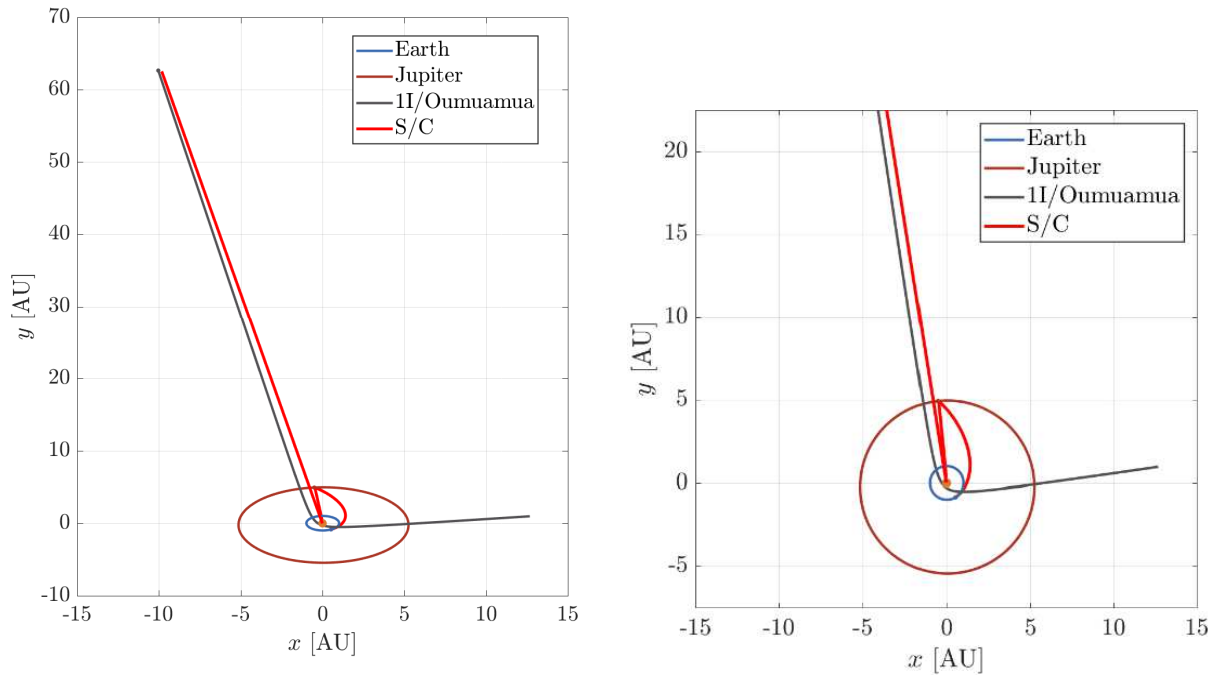

*Figure 8: Trajectory with Jupiter flyby and solar Oberth maneuver*

The minimum DeltaV budget for the year 2021 is decomposed in Table 1. The $C_3$ after Earth escape is 99.7 km²/s².

*Table 1: DeltaV budget for minimum DeltaV trajectory in 2021.*

| Maneuver | DeltaV [m/s] |
|---|---|
| Earth escape | 9985 |
| Jupiter flyby | 4166 |
| Solar Oberth maneuver | 4181 |
| ***Total DeltaV*** | ***18332*** |

The duration of the individual trajectory legs is shown in Table 2. The Earth – Jupiter outbound leg has a duration of about 18 months, about 10.5 months for getting from Jupiter to the Sun, and 69 months from the Sun to 1I/'Oumuamua.

*Table 2: Start and end dates of the trajectory legs for a mission using Jupiter flyby and Oberth maneuver.*

| Trajectory leg | Start date | End date |
|---|---|---|
| Earth to Jupiter | 2021 April 30 | 2022 October 6 |
| Jupiter to Sun | 2022 October 6 | 2023 July 14 |
| Sun to 1I/'Oumuamua | 2023 July 14 | 2029 April 23 |

The corresponding mass budget for the SLS and Falcon Heavy can be found in Table 3 and Table 4 respectively. We assume an up-scaled version of the Star solid propellant engine for both the powered flyby at Jupiter and the solar Oberth Maneuver. A total mass of 6000 kg is assumed after Earth escape for SLS and 1800 kg for Falcon Heavy. We estimate the shield mass, using values from the Solar Parker probe, and multiply the combined wet mass of the solid rocket booster and the final mass after solid booster ejection by 4.4%.

*Table 3: Propulsion characteristics and mass budget for Jupiter flyby plus Oberth maneuver, SLS launcher*

| | Jupiter flyby | Solar Oberth |
|---|---:|---:|
| DeltaV [m/s] | 4166 | 4181 |
| Isp [s] | 292 | 292 |
| Mass ratio | 4.3 | 4.3 |
| Initial mass | 6000 | 987 |
| Final mass [kg] | 1401 | 229 |
| Propellant mass [kg] | 4599 | 758 |
| Solid rocket booster dry mass [kg] (9%) | 414 | 68 |
| Solid rocket booster wet mass [kg] | 5013 | 826 |
| Final mass after Oberth maneuver and booster ejection [kg] | | 161 |
| Heat shield mass [kg] | | 39 |
| **Spacecraft mass [kg]** | | **122** |

Table 4: Propulsion characteristics and mass budget for Jupiter flyby plus Oberth maneuver, Falcon Heavy

|  | Jupiter flyby | Solar Oberth |
|---|---|---|
| DeltaV [m/s] | 4166 | 4181 |
| Isp [s] | 292 | 292 |
| Mass ratio | 4.3 | 4.3 |
| Initial mass | 1800 | 296 |
| Final mass [kg] | 420 | 69 |
| Propellant mass [kg] | 1380 | 227 |
| Solid rocket booster dry mass [kg] (9%) | 124 | 20 |
| Solid rocket booster wet mass [kg] | 1504 | 248 |
| Final mass after Oberth maneuver and booster ejection [kg] |  | 48 |
| Heat shield mass [kg] |  | 12 |
| **Spacecraft mass [kg]** |  | **37** |

A further possibility for reducing the total DeltaV requirement is to introduce a Saturn powered flyby after the Oberth maneuver. The Saturn gravity assist provides an additional deflection of the trajectory towards 1I/Oumuamua, thereby relaxing the requirement for the Jupiter powered flyby and the Oberth maneuver. An example for this mission configuration is shown in Figure 9 with a launch date in 2020.

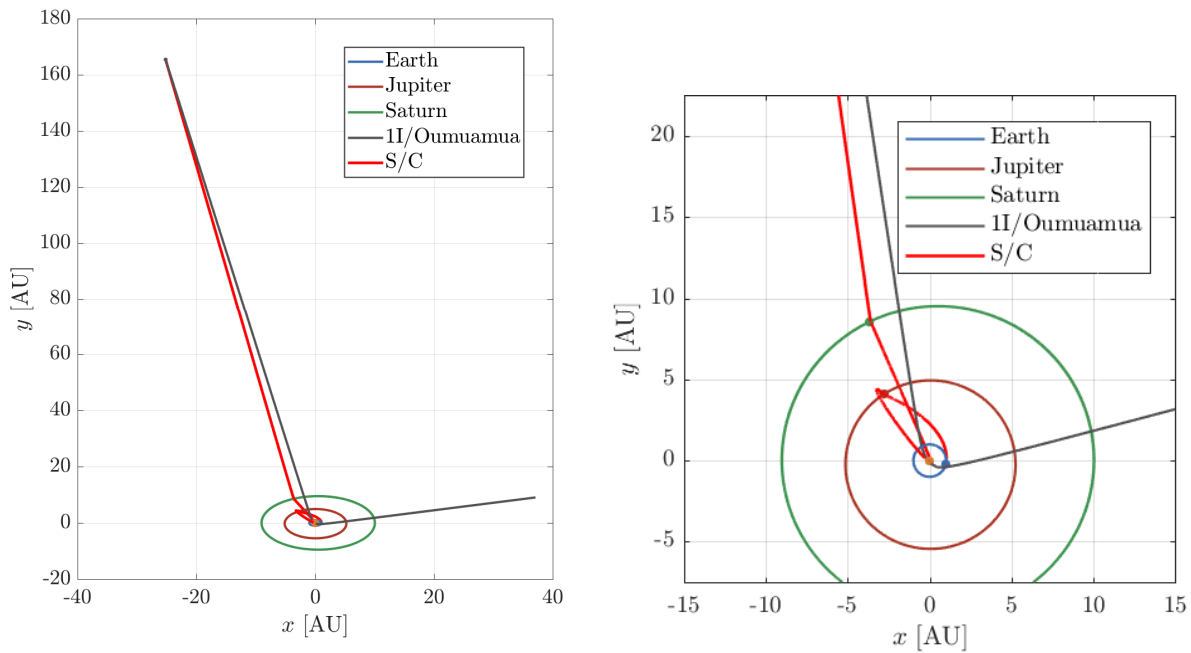

Figure 10: Trajectory with Jupiter flyby, solar Oberth maneuver and Saturn flyby.

As summarized in Table 5, the velocity requirement for this mission are listed, whereas Table 6 shows the start and end date of each individual trajectory leg. Of course due to the additional flyby, the flight duration is increased and the probe arrives at the target in 2047, at a distance of 183 AU from the Sun. Compared to the mission scenario utilizing no Saturn gravity assist, the larger distance from the Sun implies lower light intensity during payload operations, which would significantly restrict any optical measurements.

*Table 5: DeltaV budget for minimum DeltaV trajectory in 2020 using Jupiter flyby, Oberth maneuver and Saturn flyby.*

| Maneuver | DeltaV [m/s] |
|---|---|
| Earth escape | 9991 |
| Jupiter flyby | 1965 |
| Solar Oberth maneuver | 1650 |
| Saturn flyby | 0 |
| ***Total DeltaV*** | ***13606*** |

*Table 6: Start and end dates of the trajectory legs for a mission using Jupiter flyby, Oberth maneuver and Saturn flyby.*

| Trajectory leg | Start date | End date |
|---|---|---|
| Earth to Jupiter | 2021 April 30 | 2022 October 5 |
| Jupiter to Sun | 2022 October 5 | 2024 August 14 |
| Sun to Saturn | 2023 November 20 | 2025 January 26 |
| Saturn to 1I/'Oumuamua | 2025 January 26 | 2049 September 7 |

The mass budget, using two solid rocket boosters for the Jupiter powered flyby and the solar Oberth maneuver each, is shown in Table 7 for SLS and Table 8 for Falcon Heavy, respectively. Roughly 11 times higher spacecraft masses can be achieved for both, compared to the case without Saturn flyby, due to the much lower DeltaV.

*Table 7: Propulsion characteristics and mass budget for Jupiter flyby plus Oberth maneuver, SLS launcher*

|  | Jupiter flyby | Solar Oberth |
|---|---|---|
| DeltaV [m/s] | 1965 | 1650 |
| Isp [s] | 292 | 292 |
| Mass ratio | 2.0 | 1.8 |
| Initial mass | 6000 | 2753 |
| Final mass [kg] | 3021 | 1548 |
| Propellant mass [kg] | 2979 | 1205 |
| Solid rocket booster dry mass [kg] (9%) | 268 | 108 |
| Solid rocket booster wet mass [kg] | 3247 | 1314 |
| Final mass after Oberth maneuver and booster ejection [kg] |  | 1439 |
| Heat shield mass [kg] |  | 110 |
| **Spacecraft mass [kg]** |  | **1329** |

As shown in Table 8, a Falcon Heavy could sent a New Horizons-class spacecraft to 1I/'Oumuamua.

*Table 8: Propulsion characteristics and mass budget for Jupiter flyby plus Oberth maneuver, Falcon Heavy*

|  | **Jupiter flyby** | **Solar Oberth** |
|---|---|---|
| DeltaV [m/s] | 1965 | 1650 |
| Isp [s] | 292 | 292 |
| Mass ratio | 2.0 | 1.8 |
| Initial mass | 1800 | 826 |
| Final mass [kg] | 906 | 464 |
| Propellant mass [kg] | 864 | 362 |
| Solid rocket booster dry mass [kg] (9%) | 80 | 33 |
| Solid rocket booster wet mass [kg] | 974 | 394 |
| Final mass after Oberth maneuver and booster ejection [kg] |  | 432 |
| Heat shield mass [kg] |  | 33 |
| **Spacecraft mass [kg]** |  | **399** |

A Jupiter flyby would require an alignment of Jupiter with 1I/'Oumuamua, imposing constraints on potential launch dates. This is even more relevant for a combined Jupiter and Saturn flyby. Hence, one question is whether we could achieve a solar Oberth maneuver without a flyby. A sample maneuver would involve a $C_3$ of 64 km²/s² for leaving Earth and would catapult the spacecraft into an elliptic, heliocentric orbit with an aphelion of 4.1 AU. In order to perform an Oberth maneuver, a DeltaV of 7.35 km/s is required at the apoapse.

One possibility to apply the 7.35 km/s is by an electric propulsion system, such as the NASA NEXT ion engine [32,33]. Assuming a pre-Oberth mass of 5745 kg, and allowing 500 kg for ion drives, Xenon tanks, etc. results in 1224 kg of Xenon. A 10% margin results in 1350 kg, for a total mass of 7595 kg. The SLS Block 1B will be able to achieve $C_3$ = 64. A DeltaV of 7.35 km/sec over 2 years results requires an acceleration of 1.1 x $10^{-4}$ m/s² with results in a thrust of the order of 0.8 N. With the published NEXT efficiency, a total electric power of 24 kW would be needed, ideally at 4 AU. If power would be supplied by solar arrays, it results in surface area of about 1000 m² for panels with 25% efficiency. Although the Juno spacecraft uses solar arrays at about the same distance as power supply, the power generated is about two orders of magnitudes lower.

A less mature technology for supplying power to a spacecraft at 4 AU distance would be a laser electric propulsion system, such as proposed by Landis et al. [34] where power is beamed over planetary distances and converted into electricity via solar arrays. Taking a sample specific mass for the power conversion system of 0.25 kg/kW from Brophy [35], it would result in a mass of 96 kg. The beam power of the corresponding laser infrastructure would be on the order of 1 MW.

In the following section, potential near-term technologies that could be used for type of trajectory using the Earth-Jupiter-Sun-1I/Oumuamua sequence are presented.

## 3. Concepts and Technologies

As shown previously, chasing 1I/'Oumuamua with a near-term launch date (next 5-10 years), is a formidable challenge for current space systems. However, we will demonstrate that currently planned launch systems and existing technologies can be used for such a mission.

### 3.1 Technologies for Solar Oberth Maneuver

Three launch systems that will be available in the next 5 to 10 years that could be used for a mission to 1I/'Oumuamua are NASA's Space Launch System (SLS), the SpaceX Falcon Heavy, and the SpaceX Big Falcon Rocket (BFR). Nominally a single launch architecture, for example, using the SLS would simplify mission design. However other launch providers project promising capabilities in the next few years. One potential mission architecture is to make use of SpaceX's Big Falcon Rocket (BFR) and their in-space refueling technique with a launch date in 2025. Using the BFR however eliminates the need for multi-planet flybys to build up momentum for a Jupiter trajectory. Instead via direct launch from a Highly Eccentric Earth Orbit (HEEO) the probe, plus various kick-stages, is given a C3 of 100 km²/s² into an 18 month trajectory to Jupiter for a gravity assist into the solar Oberth maneuver. A multi-layer thermal shield protects the spacecraft, which is boosted by a high-thrust solid rocket stage at perihelion. The KISS Interstellar Medium study computed that a hyperbolic excess velocity of 70 km/s was possible via this technique, a value which achieves an intercept at about 85 AU in 2039 for a 2025 launch. As the calculations in section 2.2 have shown, the achievable hyperbolic excess velocity is more than enough for an earlier launch date and reaching 1I/'Oumuamua within 8 to 14 years. More modest figures can still fulfill the mission, such as 40 km/s with an intercept at 155 AU in 2051. With the high approach speed a hyper-velocity impactor to produce a gas 'puff' to sample with a mass spectrometer could be the serious option to get in-situ data.

Similar to the KISS study [29], we assume a solid rocket booster for the Oberth maneuver kick burn. However, an additional solid rocket booster is used during the Jupiter flyby and no deep space maneuver is performed. We assume an Isp of 292 s for the solid rocket booster.

According to [36], using a SLS B1 with an EUS upper stage can achieve a $C_3$ of 100 km²/s² with a 6 metric ton payload. Using this value and calculating the masses for the solid rocket boosters, the sun shield and the spacecraft wet mass yields the values in Table 9.

*Table 9: SLS B1 EUS-based spacecraft mass budget*

| Space system element | Mass [kg] |
|---|---|
| Jupiter flyby solid rocket booster (9% dry mass) | 5012 |
| Solar Oberth maneuver solid rocket booster (9% dry mass) | 826 |
| Spacecraft wet mass | 90 |
| Sun shield | 71 |
| Launcher and booster adapters | |
| **Total mass** | **6000** |

As an alternative, the Falcon Heavy is considered, as it is expected to have its maiden flight in 2018. The payload mass for a $C_3$ of 100 km²/s² is roughly 1800 kg. Calculating the masses backwards yields the values in Table 10 with a wet mass of the final spacecraft of about 21 kg.

*Table 10: Falcon Heavy-based spacecraft mass budget*

| Space system element | Mass [kg] |
|---|---|
| Jupiter flyby solid rocket booster | 1504 |
| Solar Oberth maneuver solid rocket booster | 248 |
| Spacecraft wet mass | 27 |
| Sun shield | 21 |
| Launcher and booster adapters | |
| **Total mass** | **1800** |

The above architecture emphasizes urgency, rather than advanced techniques. Using more advanced technologies, for example solar sails, laser sails, and laser electric propulsion could open up further possibilities to flyby or rendezvous with 1I/'Oumuamua. In the following, first order analyses for solar and laser sail missions are given.

### 3.2 Solar and Laser Sails

For the solar sail mission, a launch from Earth orbit is assumed, given a time to launch of 3 to 4 years. The velocity requirement is ~55 km/s, suggesting a lightness number for the mission of 0.15, and a characteristic acceleration of 0.009 m/s2. This requires a sail loading of 1 g/m², advanced materials with light payloads might achieve 0.1 g/m². Given this, for different spacecraft masses assuming a sail loading of $\sigma = 1$ g/m² sail design leads to the values shown in Table 11 for a circular and square-shaped sail.

*Table 11: Solar sail parameters with respect to spacecraft mass*

| Spacecraft mass [kg] | Sail area [m²] | Circular radius [m] | Square size [m] |
|---|---|---|---|
| 0.001 | 1 | 0.56 | 1 |
| 0.01 | 10 | 1.78 | 3 |
| 0.1 | 100 | 5.64 | 10 |
| 1 | 1000 | 17.84 | 32 |
| 10 | 10,000 | 56.42 | 100 |
| 100 | 100,000 | 178.41 | 316 |

The most appropriate and practical design would assume a launch in 4 years and a 1 kg spacecraft mass and lower.

Laser-pushed sail-based missions, based on Breakthrough Initiatives' Project Starshot technology [37–39], would use a 2.74 MW laser beam, with a total velocity increment of 55 km/s, launched in 3.5 years (2021), accelerating at 1g for 3,000s, the probe size would be about 1 gram. It would reach 1I/'Oumuamua in about 7 years. A 27.4 MW laser would allow for a 10 gram probe could be launched. Higher spacecraft masses could be achieved by using different mission architectures, lower acceleration rates, and longer mission durations. However, with such a laser beaming infrastructure in place, hundreds or even thousands of probes could be sent, as illustrated in Figure 11. Such a swarm-based or distributed architecture would allow for gathering data over a larger search volume without the limitations of a single monolithic spacecraft.

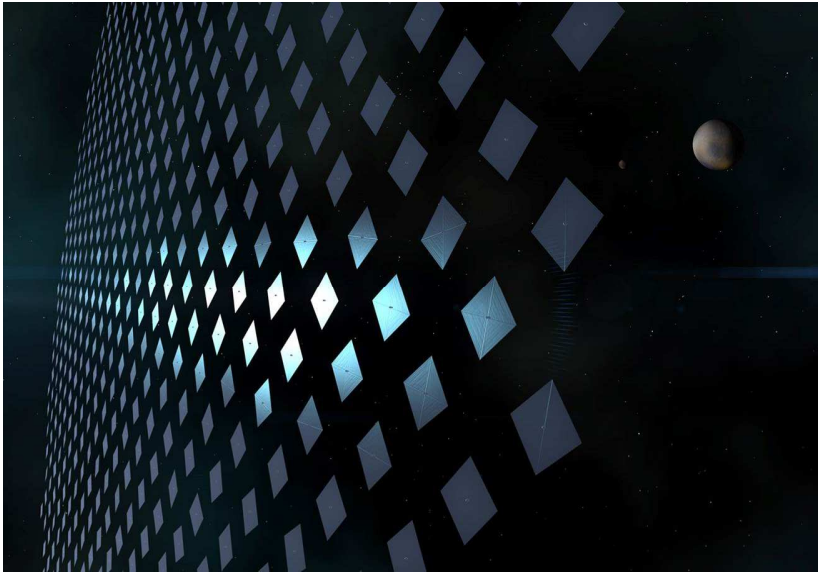

*Figure 11: Laser sail swarm (Image credit: Adrian Mann)*

An important implication is that once an operational Project Starshot beaming infrastructure has been established, even at a small scale, missions to interstellar objects flying through the solar system could be launched within short notice and could justify their development [40]. The main benefit of such an architecture would be the short response time to extraordinary opportunities. The investment would be justified by the option value of such an infrastructure.

### 3.3 Lorentz Force Acceleration

Another concept proposed by Streeman and Peck [41] is to send ChipSats into the magnetosphere of Jupiter, then using the Lorentz force to accelerating them to very high velocities of about 3,000 km/s [41–43]. However, controlling the direction of these probes might not be trivial.

### 3.4 Technologies for Deceleration

Regarding deceleration at the object, obviously existing propulsion systems could be used, e.g. electric propulsion, though limited by the low specific power of RTGs as a power source. With an intercept distance beyond the Heliosphere, into the pristine Interstellar Medium (ISM) more advanced technologies such as magnetic sails [44,45], electric sails [46], and the more recent magnetoshell braking system [47] are worth investigating. The Technological Readiness of these more advanced technologies is currently low, dependent on breakthroughs in superconducting materials manufacture, but they would multiply the scientific return by orders of magnitude.

### 3.5 Navigation

The small size of the object and its low albedo will make it difficult to observe it once it has entered deep space again. This means the navigation problem of getting a sufficiently accurate fix on 1I/'Oumuamua to get close enough to the object to send back useful data is considerable. There are therefore two alternatives for increasing the likelihood of getting sufficiently accurate remote sensing data back. Either a spacecraft with a sufficiently large aperture is sent sufficiently close to the object or a sufficiently large number of spacecraft is sent to the object with at least one spacecraft approaching close enough.

Let $m$ be the apparent optical magnitude and $H$ the absolute optical magnitude (for 1I, 22.5 - 25 depending on the optical phase. Intercept is at a distance $R$ from the Sun, and we are a distance $d$ away).

Then

$$m = H + 2.5 \log\left(\frac{d^2 R^2}{1 AU^4}\right) = H + 5 \log\left(\frac{dR}{1 AU^2}\right) \quad (4)$$

With a decent sized scope, the largest detectable m happens to be ~ *H*, which gives immediately for detection

$$d \sim \frac{1}{R} \quad (5)$$

If *R* is 100 AU, *d* is of the order of 1/100 AU, or 1.5 million km. If *R* is 150 AU, *d* is ~ 1 million km.

Due to the positional uncertainty of such a difficult-to-track object, a distributed, swarm-based mission design that is able to span a large area, should be investigated.

## 4. Conclusions

The discovery of the first interstellar object entering our solar system is an exciting event and could be a unique opportunity for in-situ observations. This article identifies key challenges of reaching 1I/'Oumuamua and ballpark figures for the mission duration and hyperbolic excess velocity with respect to the launch date. Furthermore, a more detailed mission analysis is performed for a combined powered Jupiter flyby and solar Oberth maneuver. It is demonstrated that based on currently existing technologies such as from the Parker Solar Probe, launchers such as the Falcon Heavy and Space Launch System could send spacecraft with masses ranging from dozens to hundreds of kilograms to 1I/'Oumuamua, if launched in 2021. A further increase in spacecraft mass can be achieved with an additional Saturn flyby post solar Oberth maneuver. The potential of more advanced technologies such as laser electric propulsion, solar and laser sails would also allow for chasing 1I/'Oumuamua, although their development will likely push launch dates farther into the future and might be more attractive for reaching future 'Oumuamua-like objects. The value of a laser beaming infrastructure from the Breakthrough Initiatives' Project Starshot could be the flexibility to react quickly to future unexpected events, such as sending a swarm of probes to the next object. With such an infrastructure in place today, intercept missions could have reached 1I/'Oumuamua within a year.

Future work within Project Lyra will focus on analyzing the different mission concepts and technology options in more detail and to downselect 2-3 promising concepts for further development.